\documentclass[journal,comsoc]{IEEEtran}
\usepackage[T1]{fontenc}

\ifCLASSINFOpdf
   \usepackage[pdftex]{graphicx}
  \usepackage{float}
  \usepackage{subfig}
\else
   \usepackage[dvips]{graphicx}
     \usepackage{float}
   \usepackage{subfig}
\fi
\usepackage{amsmath}
\usepackage{enumerate}
\usepackage[cmintegrals]{newtxmath}

\begin{document}
\title{A new communication paradigm: \\
    from bit accuracy  to semantic fidelity}
%
%
%
\author{Guangming~Shi,
    Dahua~Gao,
    Xiaodan~Song,\\
   Jingxuan~Chai, 
    Minxi~Yang,
    Xuemei~Xie,
    Leida~Li,
    Xuyang~Li,
\thanks{
This work has been submitted to the IEEE for possible publication. Copyright may be transferred without notice, after which this version may no longer be accessible.

Guangming Shi (corresponding author) is with Xidian University. Dahua Gao, Jingxuan Chai, Minxi Yang, Xuemei Xie and Xuyang Li are with the School of Artificial Intelligence, Xidian University.
Xiaodan Song and Leida Li are with the Guangzhou Institute of Technology, Xidian University.
Dahua Gao and Leida Li are also with the Pazhou Lab, Guangzhou 510330, China.
}
}

\maketitle

\begin{abstract}
Wireless communication has achieved great success in the past several decades. The challenge is of improving bandwidth with limited spectrum and power consumption, which however has gradually become a bottleneck with evolution going on. The intrinsic problem is that communication is modeled as a message transportation from sender to receiver and pursues for an exact message replication in Shannon's information theory, which certainly leads to large bandwidth and power requirements with data explosion. However, the goal for communication among intelligent agents, entities with intelligence including humans, is to understand the meaning or semantics underlying data, not an accurate recovery of the transmitted messages. The separate first transmission and then understanding is a waste on bandwidth. In this article, we deploy semantics to solve the spectrum and power bottleneck and propose a first understanding and then transmission framework with high semantic fidelity. We first give a brief introduction of semantics covering the definition and properties to show the insights and scope of this paper. Then the proposed communication towards semantic fidelity framework is introduced, which takes the above mentioned properties into account to further improve efficiency. Specially,  a semantic transformation is introduced to transform the input into semantic symbols. Different from the conventional transformations in signal processing area, for example discrete cosine transform, the transformation is with data loss, which is also the reason that the proposed framework can achieve large bandwidth saving with high semantic fidelity. Besides, we also discuss semantic noise and performance measurement. To evaluate the effectiveness, a case study of audio transmission is carried out. Finally, we discuss the typical applications and open challenges.
\end{abstract}


%
\IEEEpeerreviewmaketitle

\section{Introduction}
\label{intro}
%
%
%
%
\IEEEPARstart{T}{he} past few decades have witnessed an explosion on wireless data traffic due to the continuously increased connected mobile devices, enriched media content, high quality data requirement and increasing online time. To meet the continuously increased demands, wireless communication systems have evolved from the earlier 1G to the latest 5G and achieved great success. European Telecommunications Standards Institute specifies that the downlink and uplink rate should reach up to 20G/s and 10G/s in 5G scenarios. Plethora of applications including telemedicine, intelligent transportation, industrial automation, robots and remote surveillance, AR/VR and health care, will be supported with the commercial use of 5G in 2020. However, it is predicted 5G will be out of usage in 2030. The upcoming tactile internet, holographic stereo video communication and remote surgeries are all beyond the capacity of current 5G systems. It is required to go to the next change.
 
However, the advent of historical wireless technology is based on huge cost and shows the following trends:
\begin{enumerate}[1)]
	\item \textbf{The decreasing available spectrum}. In 2G/3G, spectrum with 1.6-2.0 GHz are occupied. 4G takes that with 2.0-8.0 GHz as a carrier. The waves within 3 to 300 GHz are used in the latest 5G. 6G is now considering spectrum with Terahertz and visible light for communication. From the tendency, it can be observed that the spectrum and bandwidth both goes higher and higher. The out of usage in spectrum can almost be predicted in the future.
	\item \textbf{The increasing power consumption}. With the upgrading of wireless communication,  the power consumption continues increasing due to the high working frequency, at which the signal attenuates fast with the propagation distance increasing. Either power or the network intensity is enhanced to be comparable with that working at low frequency. It is estimated that to reach the same coverage as 4G network, more than 2 trillion yuan needs to be investigated since the power consumption of 5G and the operator considers to close parts of 5G base stations in the midnight.  
\end{enumerate}
In the earlier stages, it is possible to achieve wider bandwidth on the cost of spectrum and power. With iterations going on, the cost becomes expensive and unbearable to go to the next stage if continues the trends. It is desired to go beyond the bottleneck and limitation.

The bottleneck can date back to 1948 when Shannon proposed the model of communication \cite{shannon1948a}, in which communication is modeled as a replication of the message sent by the transmitter at receiver and semantics of communication is irrelevant to the engineer problem. After that, the communication system targets at an accurate transmission of data and exact recovery of transmitted signals at receiver; it pursues for bit level fidelity. The performance is measured in bitrate and bit error. Obviously, the explosion of data will lead to a lack of available bandwidth.  

By contrast, communication among humans seems to be without such limitation. Acoustic waves with limited bandwidth are used as the carrier among humans. However, the obtained information has been increased largely comparing with our ancestors. The secret lies in the intelligence within brain, which differs conventional communication systems (CCSs) in two aspects: 
 \begin{figure}[!t]
    \centering
    \includegraphics[width=0.9\linewidth]{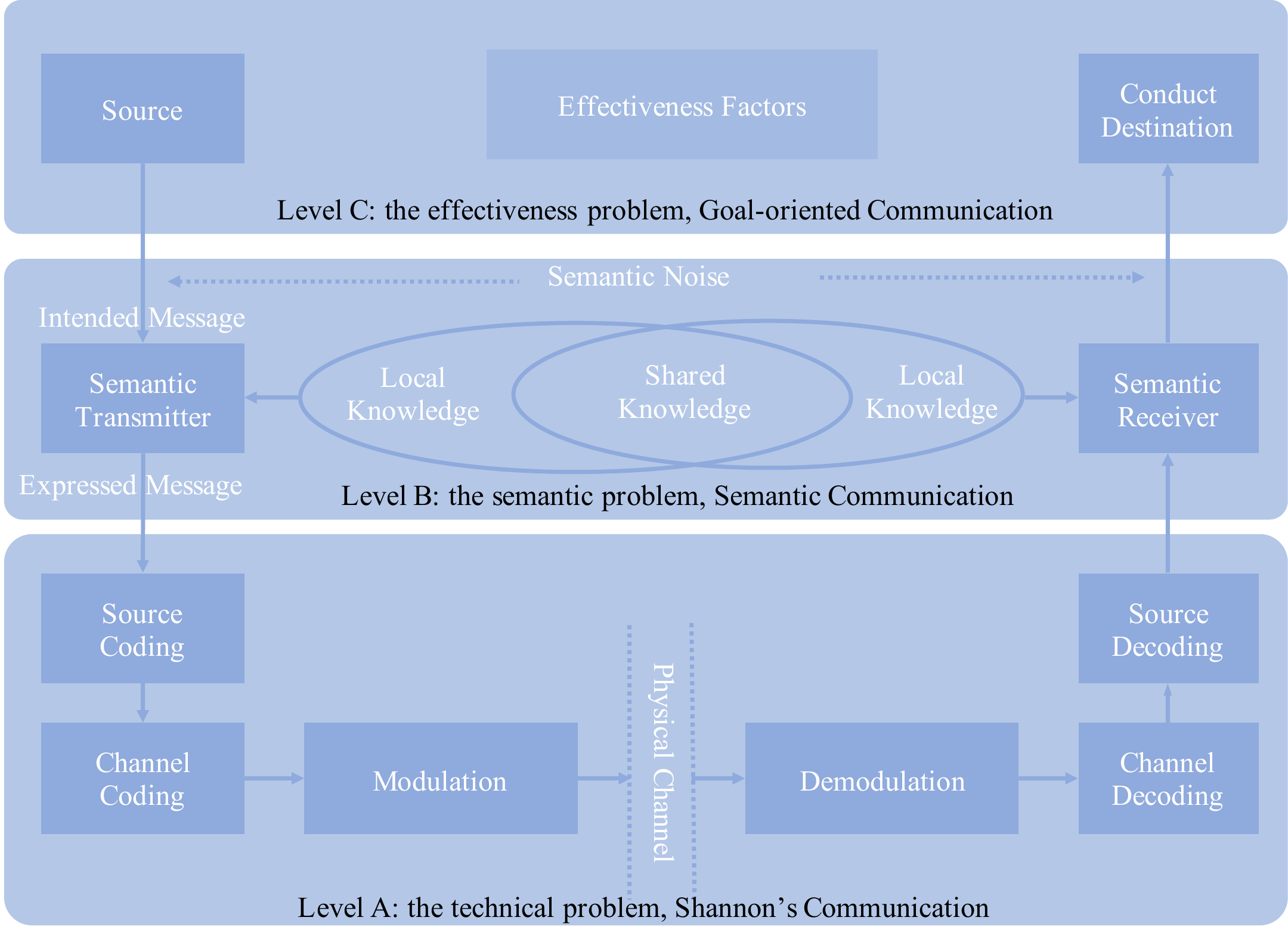}
    \caption{The three level communication model proposed by Weaver \cite{shannon1963mathematical}. The local and shared knowledge in Level B is proposed by \cite{bao2011towards}. }
    \label{weaver}
\end{figure}

\begin{enumerate}[1)]
	\item \textbf{Semantic oriented communication}. The communication among humans concerns the semantics or meaning, not the exact words and pronunciation. Communication among intelligent agents (IAs) is similar to that among humans, which takes semantic fidelity as principle and aims at understanding semantic. We argue that communication among IAs is a replication of semantics and a communication towards semantic fidelity (CTSF).  
	
	\item \textbf{ Powerful prior knowledge within the brain}. Semantics generation relies on the pre-appointed prior knowledge, which also makes communication more efficient via reducing the redundancy between transmitted messages and that in prior knowledge at receiver.
\end{enumerate}

To understand semantics,  CCSs adopt a first transmission and then understanding scheme as shown in Fig.\ref{fig_framework}(a), which is sub-optimal. CCS mainly focuses on how to reliably and efficiently send signals from sender to receiver, without considering the semantics underlying bits. It fundamentally regards the sender and receiver as agents without intelligence and ignores the goal for communication among IAs. Semantics is out of the scope of communication, the understanding of which is an extraction and abstraction process from received data. Large amounts of semantically irrelevant and redundant data are transmitted and consume large communication resources. It is desired to design a first understanding and then transmission framework to reach a joint optimization.

Besides CCSs, Weaver \cite{shannon1963mathematical} pointed out that semantic communication (SC) is the next level of communication following Shannon's information theory, which also targets at semantic fidelity. Weaver organized communication into three levels:

\textit{Level A. How accurately can the symbols of communication be transmitted? (The technical problem.)}

\textit{Level B. How precisely do the transmitted symbols convey the desired meaning? (The semantic problem.)}

\textit{Level C. How effectively does the received meaning affect conduct in the desired way? (The effectiveness problem.)} 

Shannon's information theory falls into Level A. Level B pursues an identified or a satisfactorily close approximate interpretation of semantics by receiver, compared with the intended ones by the sender. The desired conduct established by receiver is concerned compared the expected one by the sender in Level C. Weaver pointed out that Shannon's information theory is general enough to be extended to Level B via add a "semantic transmitter", "semantic receiver" and "semantic noise" to the model of level A, as shown in Fig. \ref{weaver}. 

Later, \cite{bao2011towards} introduce knowledge into the model and give a theoretical analysis. Note that the knowledge at sender and receiver may differ.  Recently, SC again attracts researchers' attention. \cite{sc2020,8792135} proposes that SC is an important part of the next generation 6G. \cite{xie2020deep} and \cite{xie2020lite} focuses on the problems of SC in natural language processing (NLP) and distributed neural network training and shows large saving on bandwidth, which show the promising future of SC. However, in most cases, semantics is implicitly represented underlying the input. The SC model is only limited to input with semantic messages, especially natural language, which limits the applications of SC.

After three ups and down, artificial intelligence (AI), especially deep learning, starts to blossom from 2012 and achieve great success in many areas. The emerging AI algorithms and computation has empowered the semantic representation. In this paper, we propose a new communication framework towards semantic fidelity with high efficiency. The core idea is that a semantic symbol representation is adopted as an interface for transmission. A reliable transmission of semantic symbols ensures the high semantic fidelity. To obtain the symbols, we propose to introduce a semantic transformation. Different from the transforms, for example, discrete cosine transform, the proposed transformation is with data-loss but with semantic fidelity, which thus leads to a large bandwidth-saving.

The rest of this paper is organized as following. The definition and properties of semantics is first introduced to limit the scope of this paper. Then we introduce the proposed framework. A case study of audio transmission towards semantic fidelity is established following. Finally, the typical applications and open problems are discussed.

\section{Semantics}
\label{introSC}
In this section, we give a brief introduction on semantics covering the definition and properties to clearly limit the scope of this paper. Though commonly used in AI, for example, semantic segmentation, semantic computing and semantic web, it is complicated to formally define and measure semantics within data/signals/message. A comprehensive survey and analysis of semantic information can be found in \cite{floridi2009philosophical}.

\subsection{Definition and Scope}
\label{def}
Semantics commonly refers to the meaning underlying data  intuitively. Data can be discrete or continuous, including signals. Shannon's information theory is such general that input with any forms, for example, random numbers, texts, audios, images and videos can be taken as input of a CCS.  We limit data to the input of any existing communication systems.  Based on data, the formulation of  semantic content, a general definition of information, proposed in \cite{floridi2009philosophical} is adopted as the definition of semantics. An instance of semantics is defined, if and only if: 
\begin{itemize}
    \item the instance consists of at least one datum; 
    \item the data are well-formed (wfd), which refers to  data are organized in a correct way according the rules of a specific system;
    \item the wfd is meaningful. "Meaningful" means that the data must comply with the meaning of the chosen system, code or language. 
\end{itemize}
Besides, we further extend "Meaningful" in this paper and define semantics as a mutual labeling between eigenfunction within the above data subset or signal and a pre-defined symbol.

From the definition, it is clear to make a distinction between data and semantics. We adopt semantic symbols to represent semantics. Thus, a semantic symbol represents a subset of data. A subset of well-formed and meaningful data can be mapped to a semantic symbol. The mapping between data and semantic symbol is a many to one process, which is also one of the intrinsic reasons that semantic representation can largely reduce the bandwidth. However, the mapping is on cost of data loss since the data cannot be exactly recovered  from a given semantic symbol. Though wfd is required, the organization is not defined in a one by one order, which means limited changes within a given subset of data will not change the semantics. Thus, different subsets may be with the same semantics and a single datum may be related to different semantics. Many topics in computer vision, for example image classification and object detection, focus on learning to model the mapping in a universal way. 
\subsection{Properties}
\label{prop}
Properties of nominal semantics are discussed in the following to show the advantages of semantic representation in improving efficiency: 
\begin{enumerate}[1)]
    \item Hierarchical structure. A single semantics can be regarded as a datum. Thus, a new semantics can be formed by several semantics conforming GDI, which is called abstraction.  A further abstraction can be established to form a more abstract semantic. Then a hierarchical structure can be formed, which represents one relationship among semantics. For example, "person" is composed of head, body, arms, hands, legs and feet. The head can be divided into eyebrows, eyes, nose, mouth, ears, and head. Secondly, an abstraction requires a well-formed organization, which means a specific spatial relationship is required. When the organization is disrupted, the represented semantics will change. For example, when the head and legs exchange spatial position, a "monster", not a "person", may be referred. The hierarchical structure makes semantic representation more efficiently via abstraction, which makes a high-level SC possible and thus saves bandwidth. 
    
    \item Extensibility and openness. The hierarchical structure can be continuously enhanced or expanded with learning and training. Through continuous learning, new high-level semantics can be formed based on existing semantics. New basic data representing unseen objects can also be added to the tree.  Besides, the data composing each semantics can also dynamically added and removed. The property may benefit the following communication via an online optimized representation.
    
      \item Multi-modality. Signals with different forms can form the same semantics; one semantic can be expressed in at least one modality. For example, the word "black" corresponds to the word black in English, a black image, or even the pronunciation of "black". These signals with different forms jointly construct the semantics of "black". When the semantics of one form is perceived, the same semantic signals of other forms can be associated, which makes it possible to achieve a cross-modality communication.
\end{enumerate}

The emerging AI empowers machines and devices with intelligence. It is imaginable that tens and thousands of devices and machines will be connected with each other. Semantics will play an important role in communication. CTSF is intrinsically a transformation from data to semantic symbols. 
\section{The proposed framework}
\label{framework}
 \begin{figure*}[!t]
     	\centering
	\includegraphics[width=1.0\linewidth]{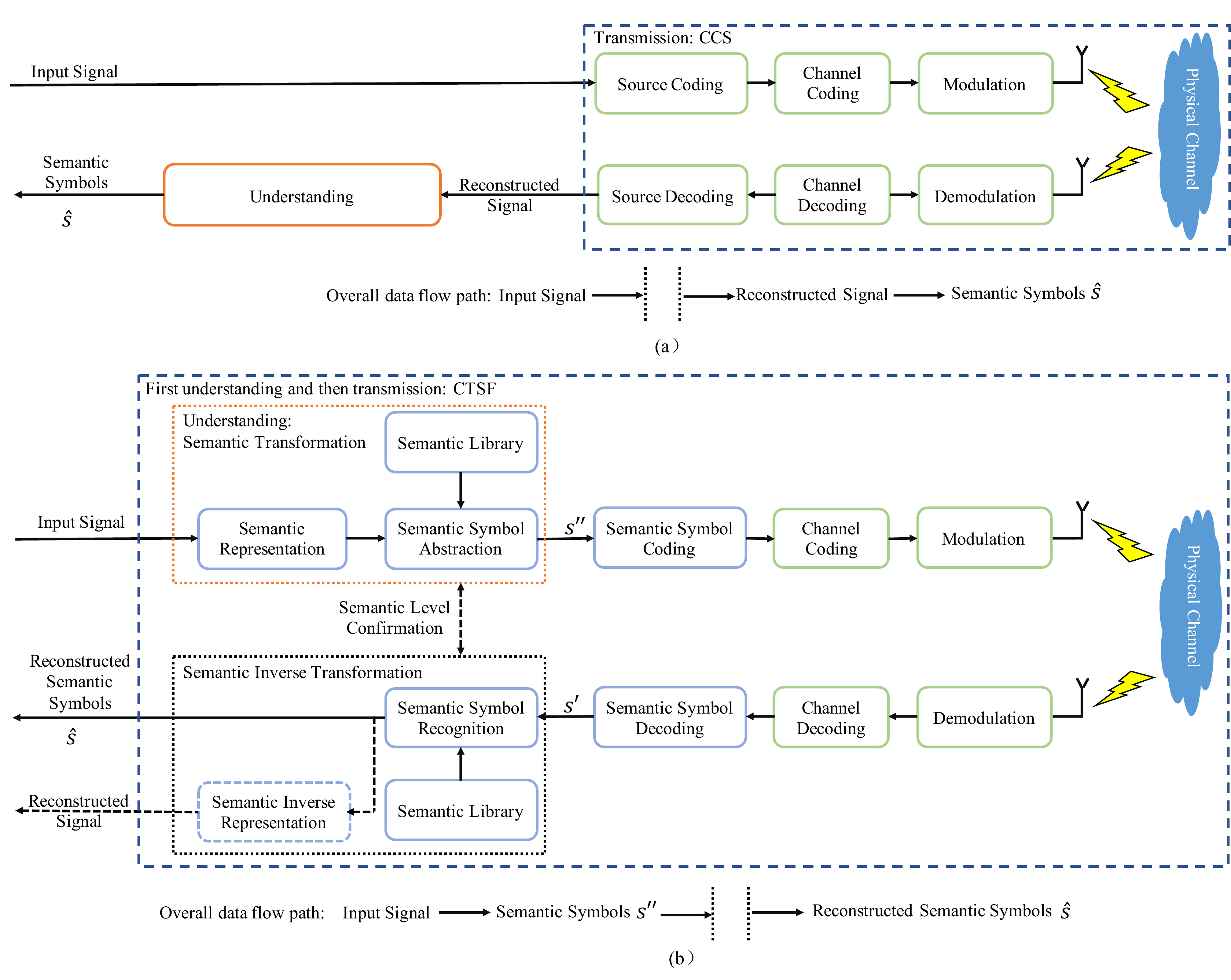}
	\caption{(a) The conventional first transmission and then understanding framework. (b) The proposed first understanding and then transmission framework denoted as CTSF. It should be noted that both $\hat{s}$ in (a) and $s''$ in (b) are not the groundth of intended semantics due to the limitation of understanding algorithms.}
	\label{fig_framework}
\end{figure*}

%
%
Fig.\ref{fig_framework}(a) and Fig.\ref{fig_framework}(b) shows the framework of the conventional systems and the proposed scheme, respectively. The conventional system adopts a first transmission and then understanding scheme. The input signal is transmitted to the receiver after source coding, channel coding and modulation. A reconstructed signal is obtained after demodulation, channel decoding, and source decoding. After that the reconstructed physical signal is converted to semantic symbols via understanding.

While the proposed scheme is a first understanding and then transmission processing as shown in Fig.\ref{fig_framework}(b). The sender transforms the input signal into semantic symbols via semantic transformation. Then the symbols are encoded into binary bits and transmitted to receiver. The transmission efficiency depends on a semantic library (SL) which is included both at sender and receiver. Via SL, an abstraction of semantic symbols can be achieved. The more capability of abstraction, the higher the efficiency is as described above. To solve the possible difference of semantic library (SL) between sender and receiver, a semantic level confirmation is introduced. In addition, we investigate how to deploy the above other properties to improve efficiency. 

The receiver is an inverse process of sender except that a semantic symbol recognition like error correction is introduced to obtain the interpreted symbols. According to requirements, signals with different modalities can be reconstructed from the interpreted semantic symbols thanks to the multi-modalities property.
 
\subsection{Semantic Transformation: from signals to semantic symbols}
\label{ST}
The semantic transformation (ST) transforms the input signal to semantic symbols with two steps: semantic representation (SR) and semantic symbol abstraction (SSA). SR is first established to convert the input to a semantic modality, for example, texts and obtain basic symbols at the lowest level of the SL. According to form of the input, algorithms in AI can be adopted, for example, automatic speech recognition (ASR) for speech. Then the basic symbols are further abstracted to reduce the amounts of transmitted symbols via SSA. Higher level abstraction represents information with more efficiency, which however requires both sender and receiver is equipped with the corresponding level of abstraction. Text summation in NLP is a typical example of SSA.

SL stores the hierarchical structure of semantics and is organized in a tree with multiple levels or depths. Each node represents a single semantic and can have multiple parent nodes and multiple child nodes. A parent node with several child nodes represent that the node is composed by the child nodes and an abstraction of the child nodes form the parent node. A node without child nodes, that is the leaf nodes, can only be that at the lowest level. SL can be constructed via offline learning, the purpose of which is to construct a knowledge database based on tasks and existing prior knowledge. It can also be manually designed, obtained via artificial intelligence algorithms, or a combination of both. Manual design refers to the construction is based on human experience. For example, for the node with semantics "human", the parent node can be man, woman, Chinese and American; the child nodes can be head, arms, body, and legs. The researches in knowledge representation, for example, knowledge graph (KG)\cite{ji2020survey} can be adopted.  Also the library can be dynamically extended via abstraction or adding new semantics online, which may benefit the following communication. Many techniques in KG and semantic computing including integration and update can be adopted to enhance the representation efficiency. For simplification, we leave it for future work.

ST can also be designed in an end to end manner, that is SR, SSA and SL are implemented implicitly. In deep-learning based image/video caption \cite{pan2017video} that converts images/videos into texts, a form of semantics, neural network is adopted as the transformation.  The transformation is learned via training with huge amounts of images/videos with labeled captions, in which the mapping between captions and training data is pre-defined in training data. 

\subsection{Semantic level confirmation}
\label{SLC}
It is common that the SL is different at sender and receiver, similar to prior knowledge is different among humans. Though the semantic symbol abstraction can improve the communication efficiency with less semantic symbols, the corresponding abstracted semantics at the receiver may not be formed. To solve the problem, a feedback, semantic level confirmation, is introduced to make sure the transmitted semantics can be understood by the receiver. For example, the sender sends "car" instead of the composed elements such as tires, and car frames. The receiver may not understand "car", that is, "car" is not within SL. Thus, the sender has to send the composed elements, and tells the receiver that these elements together represents "car".

For efficiency, a progressively probing the receiver via a feedback channel to determine which level of the SL is suggested to be used. The sender first encodes information with the highest level of semantic symbols. If the receiver cannot understand the transmitted semantic symbols, a feedback will be sent to the transmitter by the receiver and the transmitter will then transmit the semantic symbols at next lower level. Or the receiver can direct transmit which level its knowledge base is currently at. After receiving the semantic symbols can be directly understood without additional processing.

\subsection{Semantic inverse transformation: from semantic symbols to signals}
\label{SIT}
Taking the decoded semantic symbols as input, a semantic symbol recognition (SSR) and semantic symbol inverse representation (SSIR) are established sequentially. The semantic symbols $s'$ may be corrupted by channel noise and include symbols that cannot be recognized or understandable by receiver.  Error correction is implicitly achieved in SSR, which is one of the key differences with CCSs that an error correction is included in channel decoding. The problem is similar to spelling correction in NLP \cite{spellChecker}. The corpora used in training the correction model can be generated from SL. Besides, a reasoning similar to that in KG can be established to infer the sent semantic symbols based on SL when the sent ones includes unrecognized symbols. For example, "I eat a red apple" is transmitted to tell the receiver "I eat an apple". Though the receiver may not recognized "red", the fact that "I eat an apple" can be inferred via reasoning.

Usually image/videos are more vivid then abstracted texts. A reconstructed signal may be required to improve the users’ experience. A SSIR is established to reconstruct signals from the reconstructed symbols. Thanks to the multi-modalities properties, the signal can be of a or even different forms, for example, video, image, audio or text, which largely extend the applications of SC. It is exciting that recently OpenAI has claimed that a vivid image can be generated from a given text via deep learning \cite{openai_2021}. However, the reconstruction is not unique due to the many-to-one mapping.
\subsection{Semantic Noise}
\label{semanticNoise}
Semantic noise refers to the misleading between the intended semantic symbols $\boldsymbol{s}$ and received ones, which mainly comes from semantic transformation, channel noise interference, the difference between semantic libraries at the sender and receiver and semantic ambiguity. The most important one is the semantic error resulting from the mismatch between  $\boldsymbol{s}$ and $\boldsymbol{s'}$ obtained after semantic transformation, which is upper bounded by representation capability of the transformation.  Another important factor is that when channel noise is large enough, binary bits after demodulation and channel decoding may differ from the sent ones leading to symbol errors. When the SLs at sender and receiver are different, there may be symbols that cannot be recognized by receiver. Finally, when the reconstructed symbols are ambiguous, that is the same semantic symbol is used to represent multiple sets of data with different meaning, semantic noise will also be generated in SSR. 

\subsection{Performance measurement}
\label{perMea}
The performance CTSF is measured by bit-rate and semantic error rate. CTSF concerns semantic fidelity and is a replication of semantics from sender to receiver. To be compatible with the CCSs, bit rate is still adopted to measure the consumption on bandwidth. However, instead of pursuing zero bit error rate, CTSF targets at zero semantic symbol error. An easy measurement is counting for the number of difference between $\boldsymbol{s}$ and $\boldsymbol{\hat{s}}$. However, one question is that for a given signal with semantic underlying, it is difficult to obtain the groundtruth $\boldmath{s}$. Thanks to the researches in AI, a communication database may be created to guide the design of semantic transformation and system, in which $\boldsymbol{s}$ refers to the labels.  Another question is that two sequential symbols with different length may represent the same meaning. For example, “I have one apple and another apple” can be compressed into “I have two apples”. It is suggested that a measurement should be learned to measure the semantic fidelity. That is the semantic symbol error rate can be measured by $err=f(s,\hat{s})$, in which $err \in [0, 1]$.  The measurement metrics used in machine translation \cite{papineni-etal-2002-bleu} of NLP can be referred.

\section{Case study: audio transmission towards semantic fidelity}
\label{case}
\begin{figure*}[t]
    
    \centering
    \subfloat[]{
        \includegraphics[width=0.5\linewidth]{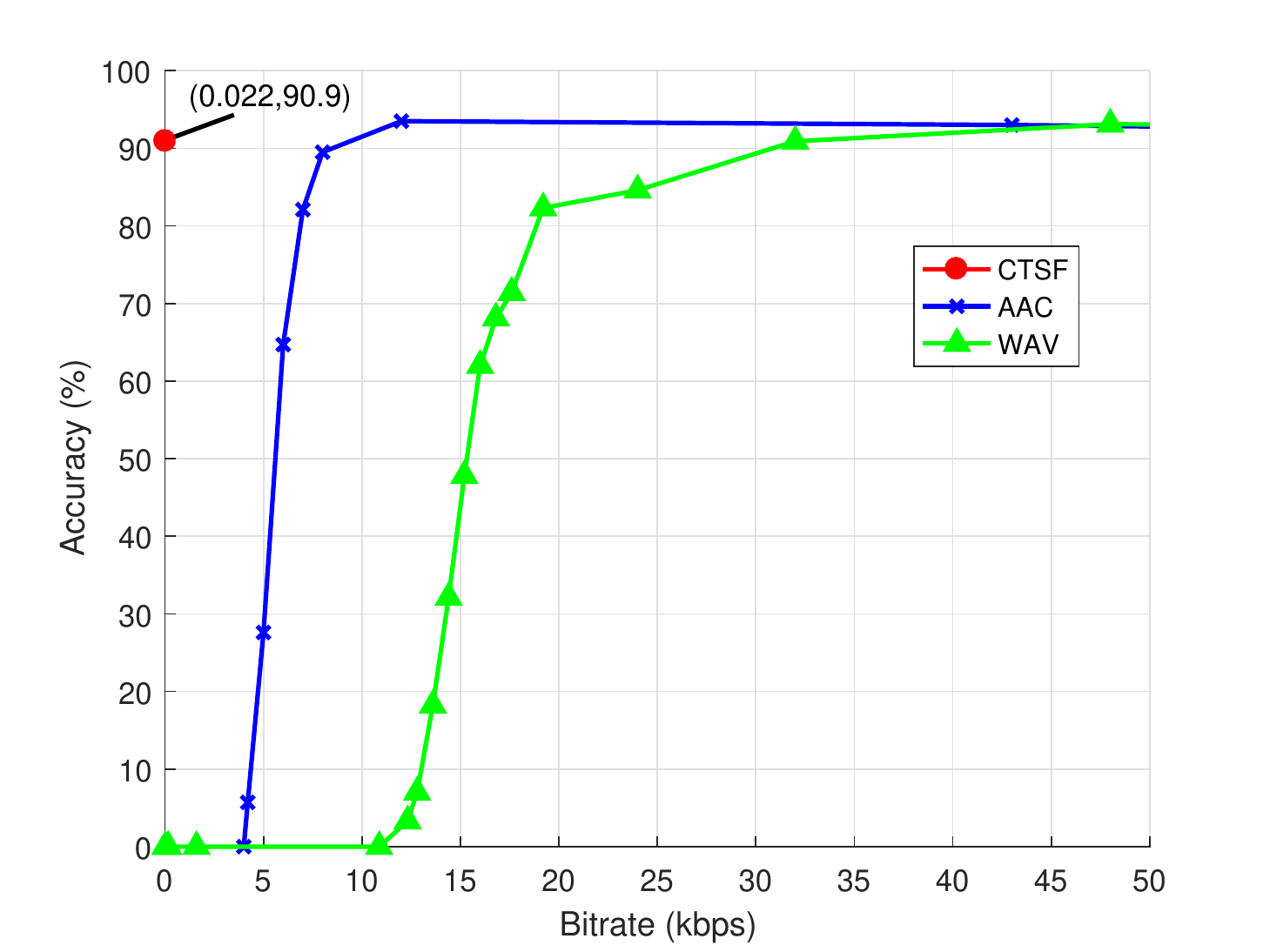} 
        \label{sc}}
    ~
    \subfloat[]{\includegraphics[width=0.5\linewidth]{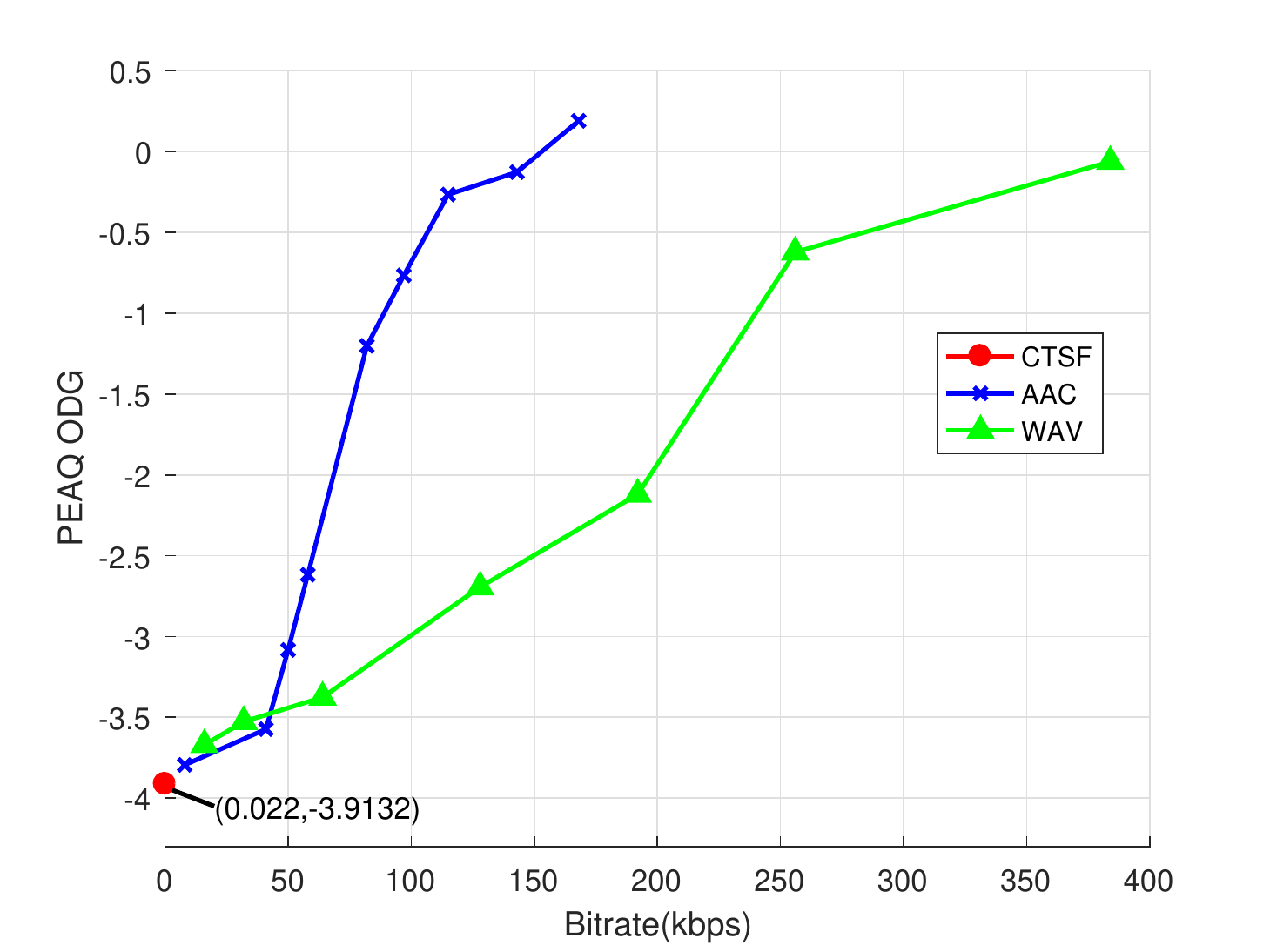}
        \label{conventional}}
    \caption{(a) The bandwidth-accuracy curve, in which semantic fidelity is measured by accuracy in Eq. (\ref{accu}). (b) The bandwidth-PEAQ ODG curve, in which PEAQ ODG measures the conventional signal fidelity.}
    \label{results}
\end{figure*}
In this section, we take audio, for example, speech, transmission as an example to illustrate CTSF over noise free channel. Two schemes are compared: CCS toward signal and bit fidelity and the proposed CTSF. It is noted the case study aims to show great potential of the proposed semantic-oriented communication in saving bandwidth and we leave the further improvement benefiting from the properties to future work.

\textbf{CCS}: The input speech is first compressed to binary bits via classical audio or speech compression algorithms. At receiver, the bits are decoded to a reconstructed speech signal. To obtain the underlying word, an automatic speech recognition (ASR) is then established. Due to compression at sender, semantic symbol errors may be within the words. The  less bandwidth available, the more severe the errors. Below a certain bandwidth, the symbols will be all corrupted by errors. With the consumed bandwidth increasing, the accuracy $1-err$ will continue to increase until all the symbols are recognized, after which, however, even more bandwidth is consumed, the fidelity cannot be improved. In summary, when measured in semantic fidelity, CCS undergoes a cliff effect. 

\textbf{CTSF}: The meaning underlying the input speech is first recognized. More specifically, the words are first extracted via ASR before encoding. Then the words are compressed into binary bits. The sent words are obtained after decompression at receiver. A post-processing may be established to recover a speech via audio synthesis and even an image or video with the same semantic can be generated from the words. The semantic fidelity depends on the difference between the groundtruth $s$ and the extracted semantic symbols $s''$.

\textbf{Experimental setup}: To avoid labeling the groundtruth of a general speech, the original input speech is generated by audio synthesis with a 16 kHz sampling rate and the semantic content $s$ is composed by a series of random integers within one to nine. The service provided by Tencent \cite{tencent} in Chinese station is adopted as ASR algorithm in both schemes. In CTSF, the recognized words are encoded by ZIP  into binary bits. In conventional communication, waveform audio file (WAV) and  advanced audio coding (AAC) are adopted to compress the original audio.  WAV losslessly compress the input video and thus can achieve the best quality but on the cost of large bandwidth. AAC is the latest lossy compression scheme for an audio and achieve large bandwidth saving compared with WAV. The recognized words from the received audio in CCS is denoted as $\hat{s}$.  Semantic fidelity is measured by
\begin{equation}
  \label{accu}
    Accuracy = 1- err =1- \frac{1}{N}\sum_{i=0}^{N}{s_i \ne \hat{s}_i},
\end{equation}
in which $\boldsymbol{\hat{s}}=\{\hat{s}_0, \hat{s}_1, ..., \hat{s}_N\}$, $\boldsymbol{s}=\{s_0, s_1, ..., s_N\}$ and $
\{s_i \ne \hat{s}_i\}=\begin{cases}
	1,&s_i \ne \hat{s}_i\\
	0,&others\\
\end{cases}
$. $N$ is set to 1000 in the experiments. A larger $Accuracy$ refers to a higher semantic fidelity. Besides,  objective difference grade (ODG) is calculated by perceptual evaluation of audio quality (PEAQ) specified in ITU BS.1387-1 to evaluate signal fidelity.  The larger PEAQ ODG, the better the quality. It is noted to evaluate PEAQ ODG of CTSF, an audio synthesis is established, though may be very different from the original audio.

\textbf{Experimental results}: The experimental results are shown in Fig.\ref{results}. It can be observed that CTSF can achieve 90.9\% accuracy at a much lower bandwidth 22 bps though with a bad PEAQ ODG lower to -4. While the accuracy of WAV and AAC at the same bandwidth is 0\%, which means the recognized words are all with errors. The proposed CTSF largely improve the semantic fidelity. To achieve the same semantic fidelity with CTSF, the required bandwidth reach up to 32kbps and 9 kbps for WAV and AAC, respectively. The bandwidth saving of CTSF reach up to 99.93\% and 99.75\%, respectively. Though AAC improves both signal and semantic fidelity, the accuracy is still far from that of CTSF. The experimental results show that the proposed CTSF can achieve high semantic fidelity with large bandwidth saving and is promising to solve bottleneck of CCSs via decreasing the required bitrate.
\section{Applications and open challenges}
\label{app}
\subsection{Applications}
\subsubsection{Ultra-low bandwidth applications}
One of the advantages of SC is the efficiency of information representation due to the high abstraction, in which a single semantic symbol can even represent a video, an image or an audio with the same semantic. It is suitable for communication with ultra-low bandwidth. Though the latest video coding standards H.266/VVC and AV1 claimed to improve efficiency by 30\%-50\%, it is impossible to achieve video communication with high fidelity via channel with ultra low bandwidth due to the large volume data of videos. SC sheds light on achieving both high quality and low bandwidth via semantic representation and powerful SL. 
\subsubsection{Holographic stereo video communication}
Holographic stereoscopic video represents information with 5D data that integrates all human senses (visual, auditory, tactile, smell, and taste) and will bring us a truly immersive remote interactive experience. It is predicted that holographic communication will replace the current remote interaction method within 10 years. Huawei reports that holographic videos obtained from multi-view cameras requires terabits per second data rate, which is, however, not supported by 5G. Although transmitting only parts of the scene that users are interested in is one solution, it is difficult to predict users' behavior. SC provides one fundamental solution. Via only transmitting semantics information and the powerful semantic library, the amount of data can be greatly reduced, which makes holographic stereoscopic video communication possible and greatly enhances the user experience.

\subsection{Open challenges}
\subsubsection{Semantic Transformation}
Semantic ambiguity \cite{ambi} is another important open challenge, especially when without context, which is also a hard problem in NLP. One difference lies in that the sender clearly know the content to be delivered to the receiver. Thus, more symbols can be transmitted to achieve disambiguation and the sender needs to remove the ambiguity as few symbols as possible. The optimal symbols to achieve the goal is still open.
\subsubsection{Semantic Information Theory}
Similar to Shannon's information theory,  it is urged to study the impact of semantics on conventional information theory, that is the reliable transmission of groundth $s$ to receiver. According the discussion of semantic noise above, the capacity depends on semantic transformation, channel noise and semantic recognition. However, a strict bound is required to show the margin of semantic communication.  Secondly, conventional communication systems assume that the transmitted symbols set are fixed. However, in semantic communication, the extensibility and openness of semantics leads to that symbols are dynamically changed. How to model the dynamic set and its impact on channel capacity is still unknown.

\section{Conclusion}
\label{conclude}
Wireless communication is intrinsically a message reproduction from sender to receiver and pursues for exact recovery, which makes the upgrade to next generation hard due to limited available spectrum and high expense. The main aim of this article is to introduce semantic communication to the wireless communication community to motivate new thoughts on conventional communication system. The proposed framework can largely alleviate the data traffic burden and improve communication efficiency via a first understanding and then communication optimization, in which semantics plays the key role. 

Besides, we'd like to attract researchers' interests in semantics, not only within communication, but also signal processing, human computer interaction and other areas. The tendency of intelligence will bring large changes to technology. How to make use of semantics and establish researches on semantics to benefit and meet new demands is a valuable and long term topic.
\section{Acknowledgments}
This work is supported by the Foundation for Innovative Research Groups of the National Natural Science Foundation of China (No. 61621005). This work is supported by Natural Science Foundation (NSF) of China (Nos. 61632019, 61836008, 61976169).
\ifCLASSOPTIONcaptionsoff
  \newpage
\fi



\bibliographystyle{./bib/IEEEtran}

%

%

\begin{IEEEbiographynophoto}{Guangming shi} [F’20] (gmshi@xidian.edu.cn) received the M.S. degree in computer control, and the Ph.D. degree in electronic information technology from Xidian University, Xi’an, China, in 1988 and 2002, respectively. His research interest includes Artificial Intelligence, Intelligent Communications, Human-Computer Interaction. He is a Professor with Xidian University. He is an IEEE Fellow and the chair of IEEE CASS Xi’an Chapter, senior member of ACM and CCF, Fellow of Chinese Institute of Electronics, and Fellow of IET. He was awarded Cheung Kong scholar Chair Professor by the ministry of education in 2012. And he won the second prize of the National Natural Science Award in 2017.
\end{IEEEbiographynophoto}

\begin{IEEEbiographynophoto}{Dahua Gao} (dhgao@xidian.edu.cn) received the Ph.D. degree in intelligent information processing from Xidian University, Shannxi, China, in 2013. He is currently a professor with the School of Artificial Intelligence, Xidian University. Dahua Gao is also with the Guangdong Artificial Intelligence and Digital Economy Laboratory, Guangzhou, Guangdong 510335, PR China.
His research interests include image processing, machine learning, and computational imaging.

\end{IEEEbiographynophoto}

\begin{IEEEbiographynophoto}{Xiaodan Song} (xdsong@xidian.edu.cn) received the B.S. and Ph.D. degrees from Xidian University, Xi’an, China, in 2012 and 2017, respectively.  Her research interests include image/video coding, multimedia communication and deep learning. She is currently a Lecture with the Guangzhou Institute of Technology, Xidian University, China.
\end{IEEEbiographynophoto}

\begin{IEEEbiographynophoto}{Jingxuan Chai} (chainorstc@gmail.com) received the B.S. degree in Communication Engineering from Xidian University in 2019. He is currently working toward the Ph.D. degree in Control Science and Engineering with the School of Artificial Intelligence, Xidian University. His research interests include pattern recognition, graph neural networks, and wireless communications.
\end{IEEEbiographynophoto}

\begin{IEEEbiographynophoto}{Minxi Yang} (mxyang@stu.xidian.edu.cn) received his B.S. degree from the School of Electronic Engineering, Xidian University, Xi’an, China, in 2019. He is currently working towards the Ph.D. degree in computer science at the School of Artificial Intelligence, Xidian University, Xi’an, China. His research interests include computer vision and representation learning.
\end{IEEEbiographynophoto}

\begin{IEEEbiographynophoto}{Xuemei Xie} (xmxie@mail.xidian.edu.cn) received the M.S. degree in Electronic Engineering from Xidian University and the Ph.D. degree in Electrical \& Electronic Engineering from the University of Hong Kong in 1994 and 2004, respectively. Her research interests are human action recognition, object detection, scene understanding, video analysis, deep learning and feature representation. She is a Professor with the School of Artificial Intelligence, Xidian University.
\end{IEEEbiographynophoto}

\begin{IEEEbiographynophoto}{Leida Li} [M'14] (ldli@xidian.edu.cn) received the B.S. and Ph.D. degrees from Xidian University, Xi’an, China, in 2004 and 2009, respectively. His research interests include multimedia quality assessment, affective computing, information hiding, and image forensics. He is currently a Professor with the Guangzhou Institute of Technology, Xidian University, China. He has served as an SPC for IJCAI 2019-2021, the Session Chair for ICMR in 2019 and PCM in 2015, and the TPC for AAAI in 2019, ACM MM 2019-2020, ACM MM-Asia in 2019, ACII in 2019, and PCM in 2016. He is currently an Associate Editor of the Journal of Visual Communication and Image Representation and the EURASIP Journal on Image and Video Processing.
\end{IEEEbiographynophoto}

\begin{IEEEbiographynophoto}{Xuyang Li} (xyLee@stu.xidian.edu.cn) received the M.S. degree from Xidian University, Xi'an, China in 2014. He is currently pursuing the Ph.D. degree of Computer Science and Technology at the School of Artificial Intelligence, Xidian University. His research interests include meta-learning, deep reinforcement learning based robot intelligent interactions, communications and computer vision.
\end{IEEEbiographynophoto}





\end{document}